\documentclass[fleqn,10pt]{wlscirep}
\usepackage[utf8]{inputenc}
\usepackage[T1]{fontenc}
\usepackage{ulem}
\usepackage{comment}
\usepackage{caption} % justify captions
\usepackage{xcolor}
\usepackage{multirow}
\usepackage{multicol}
\usepackage{mathtools}
\usepackage[small]{subfigure}
\usepackage{siunitx}
\usepackage{orcidlink}
\usepackage{pifont}
\newcommand{\cmark}{\ding{51}}
\newcommand\crule[1][black]{\textcolor{#1}{\rule{0.15cm}{0.2cm}}}

\definecolor{GREEN}{RGB}{0, 140, 80}

\makeatletter
\newcommand*{\starsection}[1]{%
  \section*{#1}%
  \NR@gettitle{#1}%
}
\makeatother
\makeatletter
\newcommand*{\starsubsection}[1]{%
  \subsection*{#1}%
  \NR@gettitle{#1}%
}
\makeatother
\makeatletter
\newcommand*{\starsubsubsection}[1]{%
  \subsubsection*{#1}%
  \NR@gettitle{#1}%
}
\makeatother

\newcommand{\figlabel}[1]{Figure~\ref{#1}}
\newcommand{\figabvlabel}[1]{Fig.~\ref{#1}}
\newcommand{\seclabel}[1]{\nameref{#1}}
\newcommand{\tablabel}[1]{Table~\ref{#1}}

\definecolor{Wine}{RGB}{150,0,90}
\definecolor{DarkBlue}{RGB}{17,0,188}
\definecolor{Blue}{RGB}{0,19,243}
\definecolor{LightBlue}{RGB}{0,110,255}
\definecolor{Teal}{RGB}{24,209,197}
\definecolor{Green}{RGB}{92,255,83}
\definecolor{Lime}{RGB}{186,248,0}
\definecolor{Yellow}{RGB}{255,207,0}
\definecolor{Orange}{RGB}{255,99,0}
\definecolor{Red}{RGB}{255,0,0}

\title{Hybrid Approach for Solving Real-World Bin Packing Problem Instances Using Quantum Annealers}

\author[1]{Sebastián V. Romero\,\orcidlink{0000-0002-4675-4452}}
\author[1,*]{Eneko Osaba\,\orcidlink{0000-0001-7863-9910}}
\author[1]{Esther Villar-Rodriguez\,\orcidlink{0000-0003-3343-3737}}
\author[1,2]{Izaskun Oregi\,\orcidlink{0000-0002-3950-1668}}
\author[1]{Yue Ban\,\orcidlink{0000-0003-1764-4470}}

\affil[1]{TECNALIA, Basque Research and Technology Alliance (BRTA), 48160 Derio, Spain}
\affil[2]{EUNEIZ, 01013 Vitoria-Gasteiz, Spain}

\affil[*]{eneko.osaba@tecnalia.com}

\keywords{3dBPP, Bin Packing Problem, Logistics, Optimization, Quantum Computing, Quantum Annealer, D-Wave}

\begin{abstract}
Efficient packing of items into bins is a common daily task. Known as Bin Packing Problem, it has been intensively studied in the field of artificial intelligence, thanks to the wide interest from industry and logistics. Since decades, many variants have been proposed, with the three-dimensional Bin Packing Problem as the closest one to real-world use cases. We introduce a hybrid quantum-classical framework for solving real-world three-dimensional Bin Packing Problems (\texttt{Q4RealBPP}), considering different realistic characteristics, such as: \textit{i)} package and bin dimensions, \textit{ii)} overweight restrictions, \textit{iii)} affinities among item categories and \textit{iv)} preferences for item ordering. \texttt{Q4RealBPP} permits the solving of real-world oriented instances of 3dBPP, contemplating restrictions well appreciated by industrial and logistics sectors.
\end{abstract}
\begin{document}

\flushbottom
\maketitle

\thispagestyle{empty}

\starsection{Introduction}

The optimization on the packaging of products into a finite number of containers is a crucial daily task in the field of production and distribution. Depending on the characteristics of both packages and containers, multiple packaging problems can be formulated, generally known as Bin Packing Problems (BPP) \cite{garey1981approximation}. Within this category, the one-dimensional BPP (1dBPP) is considered as the simplest one \cite{munien2021metaheuristic}, whose goal is to pack all items into as few containers as possible. Many variants with a variable number of constraints have been proposed to deal with real situations in logistics and industry \cite{delorme2016bin}. The three-dimensional BPP (3dBPP)\cite{martello2000three}, in which each package has three dimensions: height, width and depth, is the best-known and the most challenging variant. Highlighted in several studies\cite{lodi2002heuristic,yang2010hybrid,parreno2010hybrid}, 3dBPP has a practical interest in many industrial settings. In recent years, it has been formulated to possess diverse and practical applications such as pallet loading \cite{elhedhli2019three}, road transportation \cite{ramos2018new}, air cargo \cite{paquay2016mixed}, etc. Due to its complexity, 3dBPP is also recurrently employed as a benchmark for testing newly developed methods and mechanisms \cite{paquay2018mip,silva2019exact}.

On another front, quantum computing is still at its early stage but has gathered a lot of attention from the scientific community as it offers researchers and practitioners a revolutionary paradigm for tackling different kinds of practical optimization problems \cite{lucas2014,gill2022quantum,metalearning-quantum-optimization,timeoptimal-quantum-optimization}. In particular, quantum annealers have been recently applied to a wide variety of optimization problems inspired by the fields from industry \cite{luckow2021quantum}, logistics \cite{osaba2022systematic} and economics \cite{orus2019quantum}. However, the research on BPP carried out in the quantum community is still scarce, even though BPP has been widely studied classically as an optimization problem.

The pioneering work on BPP in the field of quantum computing presents a hybrid quantum-classical method for solving the 1dBPP \cite{10.1145/3520304.3533986}, whose solver is composed of two modules: \textit{i)} a quantum subroutine with which to search a set of feasible configurations to fill one single bin and \textit{ii)} a classical computational heuristic which builds complete solutions employing the subsets given by the quantum subroutine. To deepen the performance of the quantum subroutine developed, further tests were conducted against a random sampling and a random walk-based heuristic\cite{garcia2022comparative}. Besides those two papers, an additional study formulates an atomic energy industry related problem as a 1dBPP, solving it using the D-Wave quantum annealer\cite{bozhedarov2023quantum}. Another works show \textit{quantum-inspired evolutionary computation} techniques as an alternative to tackle BPP related problems \cite{layeb2012novel,zendaoui2016adaptive,layeb2014novel}. Quantum-inspired techniques are a specific class of evolutionary algorithms which make use of quantum physics to define their operations and are designed to be executed on a classical computer\cite{zhang2011quantum}. Thus, they can not be executed on any quantum machine.

In contrast to 1dBPP, tackling 3dBPP in the quantum domain is much more challenging due to two related grounds: \textit{i)} its complexity, which increases as constraints from the real-world are taken into account and \textit{ii)} the incipient state of development of the current commercial quantum computers with capacities still limited by decoherence and errors, which could be an obstacle to solve highly-constrained problems. In this paper, we present a hybrid quantum-classical computing framework for the real-world oriented 3dBPP, which is coined as Quantum for Real Bin Packing Problem (\texttt{Q4RealBPP}). The proposed framework resorts to the the Leap Constrained Quadratic Model (CQM) Hybrid Solver (\texttt{LeapCQMHybrid}\cite{leapCQM}) of D-Wave. At the same time, \texttt{Q4RealBPP} is built on an existing code\cite{3dBPP}. This reference code is an excellent starting point, which paved the way for  these two main contributions developed in this this work.%
\noindent\begin{itemize}%
    \item\textit{Efficiency of the code}: in the reference code, the problem is formulated such that lots of variables (thus qubits) are needed. This issue meets the problem of feasibility in the context of quantum hardware in the noisy intermediate-scale quantum (NISQ) era\cite{Preskill2018}. Therefore, the optimization of the code is crucial for dealing with complex problems. \texttt{Q4RealBPP} provides a step forward on this aspect by exploring how the constraints coined as \textit{Intrinsic restrictions} can be polished. These restrictions are the ones needed for the BPP definition (e.g. do not place cases outside a bin), and they were previously introduced in the reference code. Thus, the novel work performed on this specific facet has involved, among others, the redefinition of some mathematical formulas that define these intrinsic restrictions, and the elimination of a redundant optimization objective. Thanks to this procedure, problems can be formulated by using fewer variables, which directly impacts the capacity of the framework to address larger instances.
    \item\textit{Applicability of the tool}: the reference code is oriented toward solving the most basic variant of the 3dBPP by only considering the dimension of both packages and bins. This setting falls far from real client demands, where other features such as weights, load balancing or incompatibilities,among others, are likely to take part. We have elaborated on this direction by implementing a set of constraints named \textit{Real-World BPP Restrictions}. All these constraints represent an added value for \texttt{Q4RealBPP}, having required a significant extension of the mathematical formulation of the problem. We deepen in this aspect in following sections.%
    %and the ones relevant from a real-world perspective (e.g. do not store incompatible items into the same bin)
\end{itemize}%

Taking these factors into account, \texttt{Q4RealBPP} is oriented to industrial and logistics related fields, contemplating problems such as the organization of port containers, the introduction of packages in delivery vans and trucks or the placement of foodstuffs on distribution pallets, among others. With the aid of a hybrid quantum-classical method, \texttt{Q4RealBPP} represents a solid step forward to solve 3dBPP with the clear purpose of facing real-world focused problems well appreciated by final users and companies. Features contemplated by \texttt{Q4RealBPP}  involve \textit{i)} dimensions of packages and bins, \textit{ii)} maximum weight allowed per bin, \textit{iii)} positive and negative affinities among item categories and \textit{iv)} preferences for package ordering (in terms of load bearing and load balancing). To demonstrate its application, we have conducted an experimentation composed of 12 different instances serving as illustrative examples. Additionally, \texttt{Q4RealBPP} allows users to easily build flexible and well-defined instances to adapt a plethora of real-world situations to be solved in the quantum computer.

The rest of the article is organized as follows: in~\seclabel{sec:formulation} the 3dBPP formulation and its corresponding notation are presented. Moreover, a detailed study of the computational resources needed for arbitrary instances is carried out. In~\seclabel{sec:results}, the applicability of this tool is tested using a set of realistic instances as input. Finally, the conclusions led by the presented results and our future plans are given in~\seclabel{sec:conc}.

\starsection{Mathematical formulation}\label{sec:formulation}

In this section, we describe in detail the mathematical formulation of the 3dBPP variant tackled in this research. First, input parameters and variables that compose the problem are shown in~\tablabel{tab:params_vars_used}.
\noindent\begin{table}[!h]
\centering
\begin{tabular}{lp{15cm}}
\hline
\multicolumn{1}{l}{\bf Parameters} \\
$I,J,K_i,Q$ & sets of items, bins, orientations of $i$-th item (see~\figlabel{fig:orientations}) and relative positions between items (see~\figlabel{fig:positions}). \\
$m,n$ & number of items and bins. \\
$l_i,w_i,h_i,\mu_i$ & length, width, height and weight of item $i\in I$. \\
$L,W,H,M$ & length, width, height and maximum capacity (\textit{optional}) of bins. \\
$A^\text{pos},A^\text{neg}$ & sets of positive and negative affinities (incompatibilities) between items of type $\alpha$ and $\beta$ (\textit{optional}). \\
$\eta$ & maximum mass ratio between two items where one is placed at the top of the other with $\eta>1$ (\textit{optional}). \\
$(\tilde{L},\tilde{W})$ & target center of mass of the resultant packings (\textit{optional}). \\ [2mm]
\multicolumn{1}{l}{\bf Variables} \\
$v_j$ & binary variable that represents if bin $j\in J$ is used. \\
$u_{i,j}$ & binary variable that represents if item $i$ is added to bin $j$. \\
$r_{i,k}$ & binary variable that represents if the orientation $k\in K_i$ is applied to the item $i$. They are used to compute the effective length, width and height $(x'_i,y'_i,z'_i)$ of the $i$-th item (see~\eqref{eq:x_prime}-\eqref{eq:z_prime}). \\
$x_i,y_i,z_i$ & continuous variables that return the location of the back lower left corner of item $i$ along $x$, $y$ and $z$ axes. \\
$\tilde{x}_i,\tilde{y}_i$ & continuous variables that account the relative distance between item $i$ and $(\tilde{L},\tilde{W})$ along $x$ and $y$ axes. Both are used if $(\tilde{L},\tilde{W})$ is defined previously. \\
$b_{i,k,q}$ & binary variable that returns the relative position $q\in Q$ between items $i,k\in I$. See~\figlabel{fig:positions}. \\ \hline
\end{tabular}
\caption{Parameters and variables used in our formulation.}\label{tab:params_vars_used}
\end{table}%

\starsubsection{Objectives}\label{sec:obj}

The 3dBPP can be solved as an optimization problem where a suitable cost function to minimize must be defined. In our case, this cost function is represented as the sum of three objectives. The strength given to each objective, i.e. the relevance accounted for each one, is up to the user preferences just by multiplying each objective with a suitable weight. Thus, the problem can be stated as $\min\text{ }\sum_{i=1}^3\omega_io_i$ with $\omega_i$ the weights of each objective $o_i$. In our study we will not consider this bias, i.e. $\omega_i=1\text{ }\forall i$.

The first and main objective minimizes the total amount of bins used to locate the packages. This can be achieved by minimizing%
\begin{equation}\label{eq:o_1}
    o_1 = \sum_{j=1}^nv_j.
\end{equation}
Additionally, for ensuring that items are packed from the floor to the top of the bin, avoiding solutions with floating packages, a second objective is defined by minimizing the average height of the items for all bins%
\begin{equation}\label{eq:o_2}
    o_2 = \frac{1}{mH}\sum_{i=1}^m\left(z_i + z'_i\right).
\end{equation}
Besides these two objectives reformulated from the reference code\cite{3dBPP}, we further add a third optional objective $o_3$ to take into account the load balancing feature. This concern is particularly important when air cargo planes and sailings are the chosen conveyance\cite{dahmani2016solving,zhu2019research}, for example. In those situations, packages should be uniformly distributed around a given $xy$-coordinate inside the bin. We can tackle this by computing the so-called taxicab or Manhattan distance between items and the desired center of mass for each bin. As a result, the gaps between items are also reduced. Concerning this, the third objective to be minimized is%
\begin{equation}\label{eq:o_3}
 o_3 = \frac{1}{m}\left(\frac{1}{L}\sum_{i=1}^m \tilde{x}_i + \frac{1}{W}\sum_{i=1}^m \tilde{y}_i\right),
\end{equation}
with%
\noindent\begin{equation}\label{eq:xy_abs_vars}%
    \tilde{x}_i \coloneqq \left|\left(x_i + \frac{x_i'}{2}\right)\!\text{ mod } L -\tilde{L} \right| \quad\text{and}\quad
    \tilde{y}_i \coloneqq \left|y_i + \frac{y_i'}{2} -\tilde{W} \right| \quad\forall i\in I,
\end{equation}%
where $0\le x_i< nL$ (bins stacked horizontally) and $0\le y_i< W$ $\forall i\in I$. This objective term minimizes for each item the distance between the center of mass projection in the $xy$-plane and the $(\tilde{L},\tilde{W})$ coordinate of each bin.

The objectives above defined are subject to certain restrictions, which are essential to derive realistic solutions. The whole pool of constraints is separated into two categories: the ones intrinsic to the BPP definition (\seclabel{sec:intrinsic}), and the ones relevant from a real-world perspective (\seclabel{sec:real_restrictions}).

\starsubsection{Intrinsic restrictions}\label{sec:intrinsic}

\textbf{Item orientations}: the fact that inside a bin each item must have only one orientation can be implemented by using%
\begin{equation}\label{eq:orientation}
\sum_{k\in K_i}r_{i,k}=1\quad\forall i\in I.
\end{equation}%
\noindent\begin{figure}[!tb]
 \centering
 \subfigure[]{\includegraphics{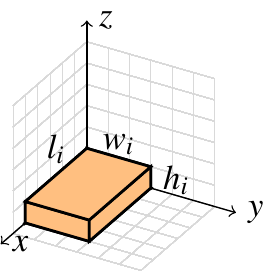}}\hspace{1mm}%
 \subfigure[]{\includegraphics{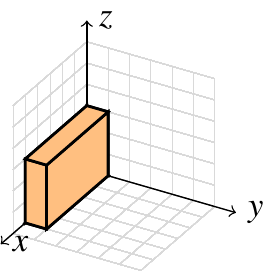}}\hspace{1mm}%
 \subfigure[]{\includegraphics{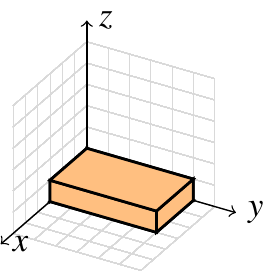}}\hspace{1mm}%
 \subfigure[]{\includegraphics{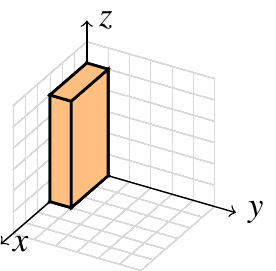}}\hspace{1mm}%
 \subfigure[]{\includegraphics{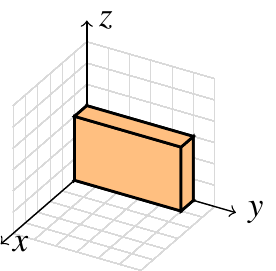}}\hspace{1mm}%
 \subfigure[]{\includegraphics{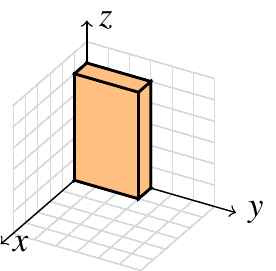}}\vspace{-3mm}%
 \caption{Set of possible orientations $k\in K_i$ for a given item $i$ of dimensions $(l_i,w_i,h_i)$. (\textbf{a}) $k = 1$, (\textbf{b}) $k = 2$, (\textbf{c}) $k = 3$, (\textbf{d}) $k = 4$, (\textbf{e}) $k = 5$, (\textbf{f}) $k = 6$. See~\tablabel{tab:unique_orientations}.}\label{fig:orientations}
\end{figure}%

Orientations give rise to the effective length, width, and height of the items along $x$, $y$ and $z$ axes %
\begin{align}
    x'_i = l_ir_{i,1} + l_ir_{i,2} + w_ir_{i,3} + w_ir_{i,4} +
    h_ir_{i,5} + h_ir_{i,6} \label{eq:x_prime} \quad\forall i\in I, \\
    y'_i = w_ir_{i,1} + h_ir_{i,2} + l_ir_{i,3} + h_ir_{i,4} +
    l_ir_{i,5} + w_ir_{i,6} \label{eq:y_prime} \quad\forall i\in I, \\
    z'_i = h_ir_{i,1} + w_ir_{i,2} + h_ir_{i,3} + l_ir_{i,4} +
    w_ir_{i,5} + l_ir_{i,6} \label{eq:z_prime} \quad\forall i\in I,
\end{align}
and because of \eqref{eq:orientation}, only one term $r_{i,k}$ is nonzero in each equation.%

It should be deemed that there could be items with geometrical symmetries, as with cubic ones where rotations do not apply. Redundant and non-redundant orientations are considered in the reference code\cite{3dBPP}. In our formulation, we previously check if these symmetries exist to define $K_i$ for each item. Thanks to this, \eqref{eq:x_prime}-\eqref{eq:z_prime} are simplified filtering out redundant orientations and leading to a formulation which uses less variables (thus qubits) to represent the same problem, where $\kappa=\sum_{i=1}^m|K_i|\le 6m$ variables $r_{i,k}$ are needed. For $i\in I_\text{c}$ with $I_\text{c}\coloneqq\{i\in I\,|\,l_i=w_i=h_i\}$ (cubic items), we can set $r_{i,1}=1$ and $0$ otherwise, thus satisfying~\eqref{eq:orientation} in advance. In~\tablabel{tab:unique_orientations}, we can see the non-redundant orientation sets for an item $i$ depending on its dimensions. This simple mechanism reduces the complexity of the problem, being favourable for the quantum hardware to implement.
\noindent\begin{table}[!ht]
\centering
 \begin{tabular}{|l|l|l|l|l|l|}
 %\cline{2-6}
 \hline
  \multicolumn{1}{|c|}{Condition} & \multicolumn{1}{|c|}{$l_i=w_i=h_i$} & \multicolumn{1}{|c|}{$w_i=h_i\neq l_i$} & \multicolumn{1}{|c|}{$l_i=h_i\neq w_i$} & \multicolumn{1}{|c|}{$l_i=w_i\neq h_i$} & \multicolumn{1}{|c|}{$l_i\neq w_i\neq h_i$} \\ \hline
  $K_i$ & $\varnothing$ & $\{1,3,4\}$ & $\{1,2,3\}$ & $\{1,2,5\}$ & $\{1,2,3,4,5,6\}$ \\ \hline
 \end{tabular}
\caption{Subsets of non-redundant orientations for item $i$ for a satisfied condition following \eqref{eq:x_prime}-\eqref{eq:z_prime}.}\label{tab:unique_orientations}
\end{table}%

\textbf{Non-overlapping restrictions}: since we are considering rigid packages, i.e. they can not overlap, a set of restrictions need to be defined to overcome these configurations. For this purpose, at least one of these situations must occur (see~\figlabel{fig:positions})%
\begin{align}\label{eq:geo0}
    &\text{Item $i$ is at the left of item $k$ ($q=1$):} &&-(2 - u_{i,j}u_{k,j}-b_{i,k,1})nL+x_i+x'_i-x_k\le0 &&\forall i,k\in I,\text{ }\forall j\in J, \\ \label{eq:geo1}
    &\text{Item $i$ is behind item $k$ ($q=2$):} &&-(2 - u_{i,j}u_{k,j}-b_{i,k,2})W+y_i+y'_i-y_k\le0 &&\forall i,k\in I,\text{ }\forall j\in J, \\ \label{eq:geo2}
    &\text{Item $i$ is below item $k$ ($q=3$):} &&-(2 - u_{i,j}u_{k,j}-b_{i,k,3})H+z_i+z'_i-z_k\le0 &&\forall i,k\in I,\text{ }\forall j\in J, \\ \label{eq:geo3}
    &\text{Item $i$ is at the right of item $k$ ($q=4$):} &&-(2 - u_{i,j}u_{k,j}-b_{i,k,4})nL+x_k+x'_k-x_i\le0 &&\forall i,k\in I,\text{ }\forall j\in J,\\ \label{eq:geo4}
    &\text{Item $i$ is in front of item $k$ ($q=5$):} &&-(2 - u_{i,j}u_{k,j}-b_{i,k,5})W+y_k+y'_k-y_i\le0 &&\forall i,k\in I,\text{ }\forall j\in J, \\ \label{eq:geo5}
    &\text{Item $i$ is above item $k$ ($q=6$):} &&-(2 - u_{i,j}u_{k,j}-b_{i,k,6})H+z_k+z'_k-z_i\le0 &&\forall i,k\in I,\text{ }\forall j\in J.
\end{align}%
As discussed with the orientation variable $r_{i,k}$ in~\eqref{eq:orientation}, the relative position between items $i$ and $k$ must be unique, so%
\begin{equation}\label{eq:case_relation}
    \sum_{q\in Q}b_{i,k,q}=1\quad\forall i,k\in I.
\end{equation}%
\noindent\begin{figure}[!b]
    \centering\vspace{-5mm}
    \subfigure[]{\includegraphics{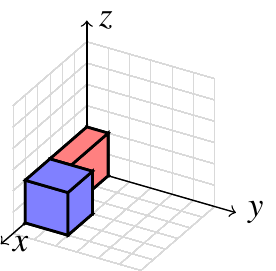}}\hspace{1mm}%
    \subfigure[]{\includegraphics{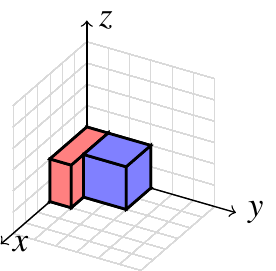}}\hspace{1mm}%
    \subfigure[]{\includegraphics{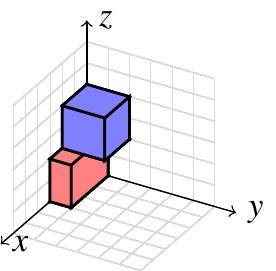}}\hspace{1mm}%
    \subfigure[]{\includegraphics{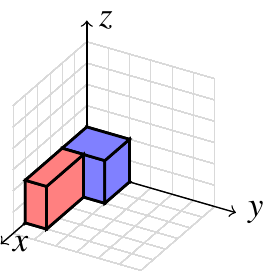}}\hspace{1mm}%
    \subfigure[]{\includegraphics{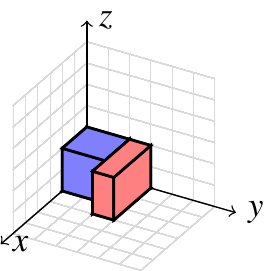}}\hspace{1mm}%
    \subfigure[]{\includegraphics{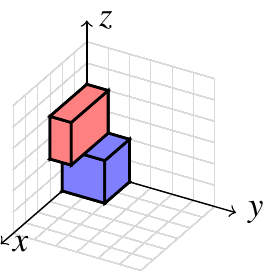}}\vspace{-3mm}%
    \caption{Representation of $b_{i,k,q}$ activated for all relative positions $q\in Q$ between items $i$ and $k$. See~\eqref{eq:geo0}-\eqref{eq:geo5}. Both are in contact but it is not mandatory. (\textbf{a}) $b_{{i},{k},1}=1$, (\textbf{b}) $b_{{i},{k},2}=1$, (\textbf{c}) $b_{{i},{k},3}=1$, (\textbf{d}) $b_{{i},{k},4}=1$, (\textbf{e}) $b_{{i},{k},5}=1$, (\textbf{f}) $b_{{i},{k},6}=1$.}\label{fig:positions}
\end{figure}%

\textbf{Item and container allocation restrictions}: the following set of restrictions guarantees an appropriate behaviour during item and bin assignment. In order to avoid packing duplicates of the same item, each item must go to exactly one bin, where%
\begin{equation}\label{eq:case_one_bin}
    \sum_{j=1}^n u_{i,j}=1\quad\forall i\in I.
\end{equation}%

The following formula verifies if items are being packed inside bins that are already in use%
\begin{equation}\label{eq:use_used_bins}
    \sum_{i=1}^m(1-v_j)u_{i,j}\le0\quad\forall j\in J,
\end{equation}%
so it activates $v_j$ if needed during packaging. Bins can be activated sequentially to avoid duplicated solutions ensuring that
\begin{equation}\label{eq:order_bins}
    v_j\ge v_{j+1}\quad\forall j\in J\text{ } | \text{ }j\neq n.
\end{equation}%

\textbf{Bin boundary constraints}: in order to contemplate bin boundaries, the following set of restrictions must be met
\vspace{-8mm}
\noindent\begin{multicols}{2}%
\noindent
\begin{equation}%
    x_i+x'_i-jL \le (1-u_{i,j})nL \quad\hfill \forall i\in I,\text{ }\forall j\in J, \label{eq:x_in_bins}
\end{equation}\vspace{-10mm}%
\begin{equation}%
    x_i-(j-1)Lu_{i,j} \ge 0 \quad\hfill \forall i\in I,\text{ }\forall j\in J\text{ }|\text{ }j>1,\label{eq:x_in_bin_j}
\end{equation}%
\begin{equation}%
    y_i+y'_i-W \le (1-u_{i,j})W \quad\hfill \forall i\in I,\text{ }\forall j\in J, \label{eq:y_in_bin_j}
\end{equation}\vspace{-10mm}%
\begin{equation}%
    z_i+z'_i-H \le (1-u_{i,j})H \quad\hfill \forall i\in I,\text{ }\forall j\in J,\label{eq:z_in_bin_j}
\end{equation}%
\end{multicols}
\noindent where~\eqref{eq:x_in_bins} guarantees that items $i$ placed inside the bin $j$ are not outside of the last bin ($n$-th bin) along the $x$ axis, \eqref{eq:x_in_bin_j} ensures that item $i$ is located inside of its corresponding bin $j$ along the $x$ axis (activated if $n>1$), \eqref{eq:y_in_bin_j} confirms that item $i$ placed inside the bin $j$ is not outside along the $y$ axis, while~\eqref{eq:z_in_bin_j} ensures that item $i$ allocated inside the bin $j$ is not outside along the $z$ axis.

\starsubsection{Real-world BPP restrictions}\label{sec:real_restrictions}

In this subsection we introduce those restrictions related with the operative perspective of the problem, i.e. the ones that consider real-world industrial situations. All of the following constraints are optional in our formulation.

\textbf{Overweight restriction}: the weight of each package and the maximum capacity of containers are common contextual data to avoid exceeding the maximum weight capacity of bins, so avoid overloaded containers. We can introduce this restriction as%
\begin{equation}\label{eq:weight_constraint}
    \sum_{i=1}^m\mu_iu_{i,j}\le M\quad\forall j\in J.
\end{equation}%
This restriction is activated if the maximum capacity $M$ is given.

\textbf{Affinities among package categories}: there are commonly preferences for separating some packages into different bins (negative affinities or incompatibilities) or, on the contrary, gathering them into the same container (positive affinities). Let us consider $I_\alpha\coloneqq\{i\in I\text{ }|\text{ }\texttt{id}\text{ of }i\text{ is equal to }\alpha\}$, i.e. $I_\alpha\subset I$ is a subset of all items labelled with \texttt{id} equal to $\alpha$. Given a set of $p$ negative affinities $A^\text{neg}\coloneqq\{(\alpha_1,\beta_1),\dots,(\alpha_p,\beta_p)\}$, then the restriction will be%
\begin{equation}\label{eq:constraint_id}
\sum_{(\alpha,\beta)\in A^\text{neg}}\,\sum_{(i_\alpha,i_\beta)\in I_\alpha\times I_\beta}\,\sum_{j=1}^nu_{i_\alpha,j}u_{i_\beta,j}=0,
\end{equation}%
To activate this restriction, a set of incompatibilities must be given. Moreover, we can satisfy in advance $\nu\coloneqq 6n\sum_{(\alpha,\beta)\in A^\text{neg}}|I_\alpha||I_\beta|$ non-overlapping constraints (see~\eqref{eq:geo0}-\eqref{eq:geo5}), leading to a simpler formulation. Conversely, given a set of positive affinities \sout{$A^{-}$} $A^\text{pos}$ as stated with $A^\text{neg}$, then the restriction will be posed such that%
\begin{equation}\label{eq:constraint_id_comp}
\sum_{(\alpha,\beta)\in A^\text{pos}}\,\sum_{(i_\alpha,i_\beta)\in I_\alpha\times I_\beta}\,\sum_{j=1}^n\left(1-u_{i_\alpha,j}u_{i_\beta,j}\right)=0,
\end{equation}%
This restriction is activated if a set of positive affinities is given. If $A^\text{pos}$ and $A^\text{neg}$ are given, then both restrictions can be introduced using just one formula adding~\eqref{eq:constraint_id} and \eqref{eq:constraint_id_comp}.%

\textbf{Preferences in relative positioning}: relative positioning of items demands that some of them must be placed in a specific position with respect other existing items. This preference allows introducing the ordering of a set of packages according to their positions with respect to the axes. Thus, this preference assists in ordering for many real cases such as: \textit{parcel delivery} (an item $i$ that has to be delivered before item $k$ will be preferably placed closer to the trunk door) or \textit{load bearing} (no heavy package should rest over flimsy packages), among others.

Regarding this preference, we can define two different perspectives to treat relative positioning:%
\noindent
\begin{itemize}
    \item Positioning to avoid ($P_q^{-}$): list of items $(i,k)$ should not be in the relative position $q\in Q$ specified. So, $b_{i,k,q}=0$ is expected, favouring configurations where the solver selects $q'\in Q$ with $q'\neq q$ for the relative positioning of items $(i,k)$.%
    \item Positioning to favour ($P_q^{+}$): list of items $(i,k)$ should be in a certain relative position $q$. Activated this preference, $b_{i,k,q}=1$ ought to hold and consequently, $b_{i,k,q'}=0 \ \forall q'\neq q$.%
\end{itemize}%

Formally, these preferences are written as%
\noindent
\begin{equation}\label{eq:rel_pos}
    P_q^{-}\coloneqq\{(i,k)\in I^2\text{ }|\text{ }i<k\text{ and }b_{i,k,q}=0\}\quad\text{and}\quad P_q^{+}\coloneqq\{(i,k)\in I^2\text{ }|\text{ }i<k\text{ and }b_{i,k,q}=1\}.
\end{equation}%

These preferences could be also treated as compulsory pre-selections. In such case, the number of variables needed would be reduced, so would the search space. If we let $\smash[t]{p^{-}=\sum_{q\in Q}|P_q^{-}|}$ and $\smash[t]{p^{+}=\sum_{q\in Q}|P_q^{+}|}$ with $\smash[t]{P^{-}_q\cap P^{+}_{q'}=\varnothing}$, based on~\eqref{eq:case_relation}, the amount of variables reduced would be given by $\smash[t]{p^{-}+6p^{+}}$. Moreover, $\smash[t]{n(p^{-}+5p^{+})}$ non-overlapping constraints (see~\eqref{eq:geo0}-\eqref{eq:geo5}) are satisfied directly and can be ignored, thus simplifying the problem. In this paper, for the sake of clarity, these preferences have been applied for load bearing purposes as hard constraints (HC), as explained in the upcoming \seclabel{sec:results}.%

\textbf{Load balancing}: to activate this restriction, a target center of mass must be given. Global positions with respect to the bin as a whole (as described in objective $o_3$ in~\eqref{eq:o_3}), are fixed using the following constraints
\noindent\begin{equation}\label{eq:constraint_xy_tilde}
    \pm\frac{1}{n}\sum_{j=1}^n\left[x_i+\frac{x_i'}{2} - n(j-1)u_{i,j}L -\tilde{L}\right] \le \tilde{x}_i \quad\text{and}\quad \pm\left(y_i+\frac{y_i'}{2} -\tilde{W}\right) \le \tilde{y}_i \quad\forall i\in I.
\end{equation}%
This feature is represented in~\figlabel{fig:abs} for $(\tilde{L},\tilde{W})=(L/2,W/2)$, whose red line shows the available $\tilde{x}_i$ and $\tilde{y}_i$ values (see~\eqref{eq:xy_abs_vars}).
\noindent\begin{figure}[!htb]
\centering
\includegraphics{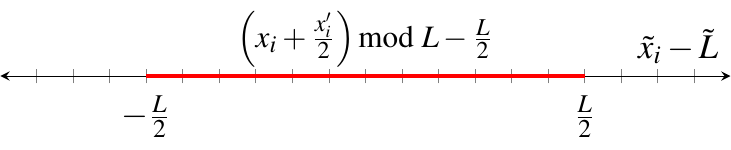}\hspace{1.5cm}%
\includegraphics{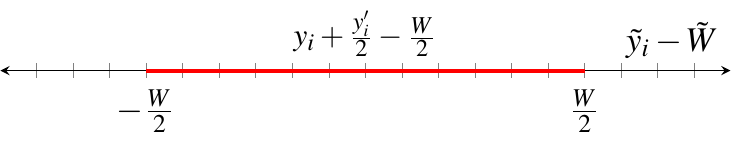}\vspace{-3mm}%
\caption{Representation of available $\tilde{x}_i$ and $\tilde{y}_i$ values ensured by the constraints given in~\eqref{eq:constraint_xy_tilde} for $(\tilde{L},\tilde{W}) = (L/2,W/2)$.}\label{fig:abs}
\end{figure}

\starsubsection{Complexity of the problem}

Regarding the complexity of the 3dBPP proposed in this research, the total amount of variables needed to tackle an arbitrary instance is given in~\tablabel{tab:amount_of_variables}, where our formulation scales as $\mathcal{O}[m^2+nm]$ in terms of variables. Additionally, the total amount of constraints required is provided in~\tablabel{tab:amount_of_constraints}, whose quantity grows quadratically as $\mathcal{O}[m^2+nm]$.%
\noindent\begin{table}[!h]
\centering
\begin{tabular}{@{}|c|r|r|r|r|r|r||r|r|@{}}
\hline
\multirow{3}{*}{\textit{Mandatory}} & \multicolumn{1}{|c|}{$n$} & \multicolumn{1}{|c|}{$x_i,y_i,z_i$} & \multicolumn{1}{|c|}{$u_{i,j}$} & \multicolumn{1}{|c|}{$v_j$} & \multicolumn{1}{|c|}{$r_{i,k}$} & \multicolumn{1}{|c||}{$b_{i,k,q}$} & \multicolumn{1}{|c|}{Binary} & \multicolumn{1}{|c|}{Continuous} \\ \cline{2-9}
& $1$ & $3m$ & - & - & $\kappa$ & $6\binom{m}{2}$ & $6\binom{m}{2}+\kappa$ & $3m $ \\ \cline{2-9}
& $\ge 2$ & $3m$ & $nm$ & $n$ & $\kappa$ & $6\binom{m}{2}$ & $6\binom{m}{2}+n(m+1)+\kappa$ & $3m$ \\ \hline
\end{tabular}\\\vspace{1mm}%
\begin{tabular}{@{}|c|r|r|r||r|r|@{}}
\hline
\multirow{3}{*}{\textit{Optional}} & \multicolumn{1}{|c|}{$n$} & \multicolumn{1}{|c|}{$\tilde{x}_i,\tilde{y}_i$} & \multicolumn{1}{|c||}{\eqref{eq:rel_pos} as HC} & \multicolumn{1}{|c|}{Binary} & \multicolumn{1}{|c|}{Continuous} \\ \cline{2-6}
& $1$ & $2m$ & $-p^{-}-6p^{+}$ & $-p^{-}-6p^{+}$ & $2m$ \\ \cline{2-6}
& $\ge 2$ & $2m$ & $-p^{-}-6p^{+}$ & $-p^{-}-6p^{+}$ & $2m$ \\ \hline
\end{tabular}
\caption{Amount of mandatory (related to~\seclabel{sec:intrinsic}) and optional (related to~\seclabel{sec:real_restrictions}) used variables depending on the number of items and bins. See~\tablabel{tab:params_vars_used}.}\label{tab:amount_of_variables}
\end{table}%
\noindent\begin{table}[!h]
\centering
\begin{tabular}{@{}|c|r|r|r|r|r||r|r|@{}}
\hline
\multirow{3}{*}{\textit{Mandatory}} & \multicolumn{1}{|c|}{$n$} & \multicolumn{1}{|c|}{\eqref{eq:orientation}} & \multicolumn{1}{|c|}{\eqref{eq:geo0}-\eqref{eq:case_relation}} & \multicolumn{1}{|c|}{\eqref{eq:case_one_bin}-\eqref{eq:order_bins}} & \multicolumn{1}{|c||}{\eqref{eq:x_in_bins}-\eqref{eq:z_in_bin_j}} & \multicolumn{1}{|c|}{Quadratic} & \multicolumn{1}{|c|}{Linear} \\ \cline{2-8}
& $1$ & $m-|I_\text{c}|$ & $7\binom{m}{2}$ & - & $3m$ & - & $7\binom{m}{2}+4m-|I_\text{c}|$ \\ \cline{2-8}
& $\ge 2$ & $m-|I_\text{c}|$ & $(6n+1)\binom{m}{2}$ & $2n+m-1$ & $(4n-1)m$ & $6n\binom{m}{2}+n$ & $\binom{m}{2}+n(4m+1)+m-1-|I_\text{c}|$ \\ \hline
\end{tabular}\\\vspace{1mm}%
\begin{tabular}{@{}|c|r|r|r|r|r||r|r|@{}}
\hline
\multirow{3}{*}{\textit{Optional}} & \multicolumn{1}{|c|}{$n$} & \multicolumn{1}{|c|}{\eqref{eq:weight_constraint}} & \multicolumn{1}{|c|}{\eqref{eq:constraint_id}/\eqref{eq:constraint_id_comp}} & \multicolumn{1}{|c|}{\eqref{eq:rel_pos} as HC} & \multicolumn{1}{|c||}{\eqref{eq:constraint_xy_tilde}} & \multicolumn{1}{|c|}{Quadratic} & \multicolumn{1}{|c|}{Linear} \\ \cline{2-8}
& $1$ & - & - & $-p^{-}-5p^{+}$ & $4m$ & - & $-p^{-}-5p^{+}+4m$ \\ \cline{2-8}
& $\ge 2$ & $n$ & $1$ & $-n(p^{-}+5p^{+})-\nu$ & $4m$ & $1-n(p^{-}+5p^{+})-\nu$ & $n+4m$ \\ \hline
\end{tabular}
\caption{Amount of mandatory (related to~\seclabel{sec:intrinsic}) and optional (related to~\seclabel{sec:real_restrictions}) considered constraints depending on the number of items and bins.}\label{tab:amount_of_constraints}
\end{table}%

\starsection{Experimental results}\label{sec:results}

In this section, we conduct an experimentation to demonstrate the applicability of \texttt{Q4RealBPP}, where the problem has been modelled as a CQM and then solved by the \texttt{LeapCQMHybrid} provided by D-Wave\cite{leapCQM}. Initially, it should be made clear that CQM refers to the mathematical model that uses quadratic objective functions and quadratic constraints on binary and integer variables. As this concept was introduced by D-Wave Systems, this company developed the hybrid solver \texttt{LeapCQMHybrid}, which also supports the definition of equality and inequality requirements. This feature brings an advantage compared to Quadratic Unconstrained Binary Optimization (QUBO), which is the native formulation for the QPUs. The CQM model and the \texttt{LeapCQMHybrid} introduce some interesting shortcuts for a more friendly use, allowing the user to provide a problem in a more intuitive and descriptive way, avoiding the translation into a mathematical formulation in the shape of a QUBO matrix.

More specifically, the \texttt{LeapCQMHybrid} workflow is as follows: first, a classical front end receives the CQM formulation of the problem as input. Then, it runs a number of parallel computation threads, where each of them are composed of two parts: a heuristic module (HM) and a quantum module (QM). On the one hand, HM tries to solve the problem by using a state of the art heuristic solver. On the other hand, QM aids this resolution by formulating different \textit{quantum queries} aiming to guide the HM toward promising regions of the search space, or finding improvements to already existing solutions. This QM performs its operations by acceding the latest available quantum computer of D-Wave. At the time of writing this paper, the most updated architecture is the \texttt{Advantage\_system}, composed of 5616 qubits organized in a Pegasus topology. For more information, we refer interested readers to the related report provided by D-Wave \cite{HybridDwave}.

Having said that, for demonstrating the applicability of \texttt{Q4RealBPP}, we have built an ad-hoc benchmark composed of 12 different instances of the 3dBPP. In order to analyze the impact of every restriction deeply, each instance is devoted to evaluate a specific feature of the problem. Also, we have generated two specific instances that bring together all the restrictions of our modelled 3dBPP. We describe in~\tablabel{tab:instances} the main characteristics of the 12 used instances.
\noindent\begin{table}[!t]
 \centering
 %\resizebox{1.0\columnwidth}{!}{
     \begin{tabular}{|l|l|c|c|c|c|c|r|r|}
      \hline
      \multicolumn{2}{|c|}{Instance} & OW & PA & INC & LB & CM & \multicolumn{1}{|c|}{Variables} & \multicolumn{1}{|c|}{Constraints} \\ \hline
      \texttt{3dBPP\_1} (51) & \figabvlabel{fig:3dBPP_1} & & & & & & 8085 & 9129 \\ \hline
      \texttt{3dBPP\_2} (51) & \figabvlabel{fig:3dBPP_1_W} & \cmark & & & & & 8189 & 17039 \\ \hline
      \texttt{3dBPP\_3} (52) & \figabvlabel{fig:3dBPP_2} & & & & & & 8406 & 9490 \\ \hline
      \texttt{3dBPP\_4} (52) & \figabvlabel{fig:3dBPP_2_fra} & & & & \cmark & & 7925 & 9009 \\ \hline
      \texttt{3dBPP\_5} (53) & \figabvlabel{fig:3dBPP_3} & & & & & & 8745 & 9858 \\ \hline
      \texttt{3dBPP\_6} (53) & \figabvlabel{fig:3dBPP_3_aff} & & & \cmark & & & 8853 & 17555 \\ \hline
      \texttt{3dBPP\_7} (46) & \figabvlabel{fig:3dBPP_4} & & & & & & 6624 & 7429 \\ \hline
      \texttt{3dBPP\_8} (46) & \figabvlabel{fig:3dBPP_4_aff} & & \cmark & \cmark & & & 6718 & 13585 \\ \hline
      \texttt{3dBPP\_9} (47) & \figabvlabel{fig:3dBPP_5_CM1} & & & & & \cmark & 7003 & 7943 \\ \hline
      \texttt{3dBPP\_10} (51) & \figabvlabel{fig:3dBPP_5_CM2} & & & & & \cmark & 8211 & 9333 \\ \hline
      \texttt{3dBPP\_11} (38) & \figabvlabel{fig:3dBPP_6_ALL1} & \cmark & \cmark & \cmark & \cmark & \cmark & 4417 & 8805 \\ \hline
      \texttt{3dBPP\_12} (38) & \figabvlabel{fig:3dBPP_6_ALL2} & \cmark & \cmark & \cmark & \cmark & \cmark & 4453 & 8973 \\\hline
     \end{tabular}
%}
\caption{Brief description of the real-world oriented restrictions activated and the amount of variables and constraints used for each instance (see Tables~\ref{tab:amount_of_variables} and~\ref{tab:amount_of_constraints}). In brackets, number of items considered. Here, OW: overweight, PA: positive affinities, INC: incompatibilities, LB: load bearing, CM: center of mass.}\label{tab:instances}
\end{table}%

The whole dataset has been generated employing an own-developed Python script (coined as \texttt{Q4RealBPP-DataGen}).  In order this paper to be as self-contained as possible, we briefly explain how \texttt{Q4RealBPP-DataGen} works. This script performs two steps to generate \texttt{Q4RealBPP}-compliant instances: firstly \texttt{Q4RealBPP-DataGen} randomly generates a defined number of items, following the package distribution and dimensions established in Ref.\cite{elhedhli2019three}. Then, the attributes regarding overweight restriction, affinities among item categories, load bearing and load balancing are completed by means of \texttt{Q4RealBPP-DataGen}. These constraints are randomly configured by the generator, except for the last two (load bearing and load balancing). These last features, as well as the bin dimensions, have been empirically set, in the search of a realistic scenario. It should be clarified that \texttt{Q4RealBPP} is a flexible framework, letting users to build their own setting not only configuring the instance of the problem but also by activating or deactivating the real-world oriented restrictions. For more information, the complete benchmark of instances and \texttt{Q4RealBPP-DataGen} are openly available in \cite{Q4RealBPPRep}. Furthermore, a deep explanation about how the dataset has been generated as well as the format of each instances is provided in \cite{osaba2023benchmark}.

In our specific use case, the \textit{preferences in relative positioning} (see~\seclabel{sec:real_restrictions}) are tested as HC for load bearing. Accounting fragility issues, one could apply the rule of choosing pairs of packages to decide on what height to place each of them based on a mass ratio $\eta$ (assuming that weight is related to fragility). Thus, defining $P_3^{-}=\{(i,k)\in I^2\text{ }|\text{ }i<k\text{ and }\mu_k/\mu_i>\eta\}$ (so $b_{i,k,3}=0\text{ }\forall(i,k)\in P_3^{-}$) and $P^{-}_6=\{(i,k)\in I^2\text{ }|\text{ }i<k,\text{ and }\mu_i/\mu_k>\eta\}$ (so $b_{i,k,6}=0\text{ }\forall(i,k)\in P_6^{-}$), this instantiation avoid configurations where items whose mass are more than $\eta$ times the mass of other ones are placed above of them.

\figlabel{fig:results} represents the results provided by \texttt{Q4RealBPP} for each of the instances described in \tablabel{tab:instances}. Regarding the running time, we have empirically determined the time it takes for the \texttt{LeapCQMHybrid} to resolve these instances to be $\SI{30}{s}$, presenting a maximum Quantum Processing Unit (QPU) access time of $\SI{0.032}{s}$ per execution. Lastly, for interested readers, all obtained results are freely available \cite{Q4RealBPPRep}.%
\noindent\begin{figure}[!tb]
	\centering
    \subfigure[Colour palette scheme used to relate \texttt{id} (item categories) with colours in the illustrative examples.\label{fig:colour_palette}]{\includegraphics{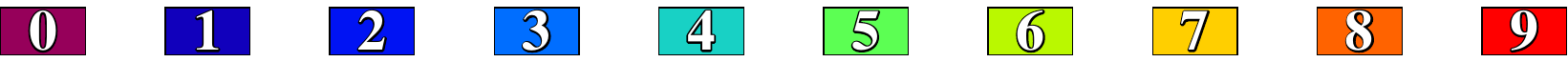}}\\%
	\subfigure[\texttt{3dBPP\_1}: 51 items without restrictions and $\sum_{i\in I}\mu_i=1776$.\label{fig:3dBPP_1}]{\includegraphics[width=\linewidth/4-1.5mm]{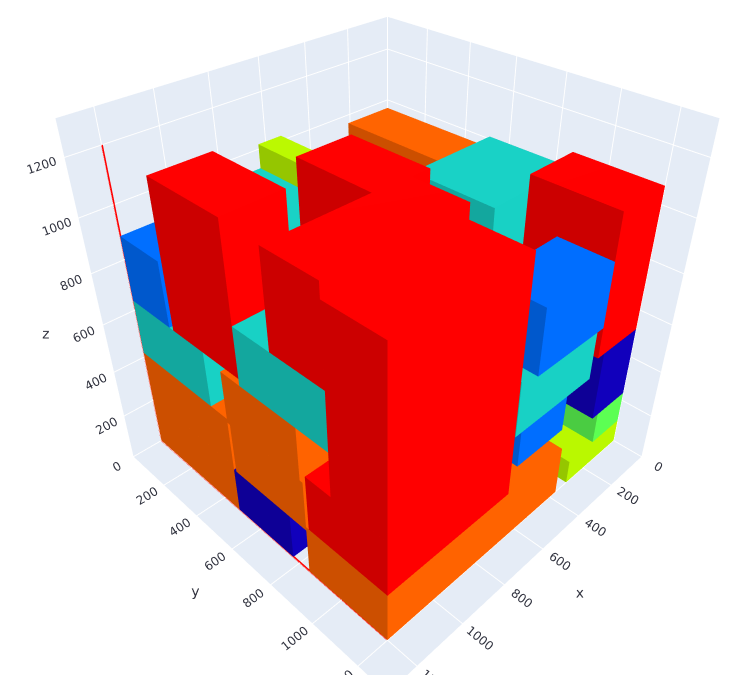}}\hspace{2mm}%
	\subfigure[\texttt{3dBPP\_2}: inst. \texttt{3dBPP\_1} with $M=1000$.\label{fig:3dBPP_1_W}]{\includegraphics[width=\linewidth/4-1.5mm]{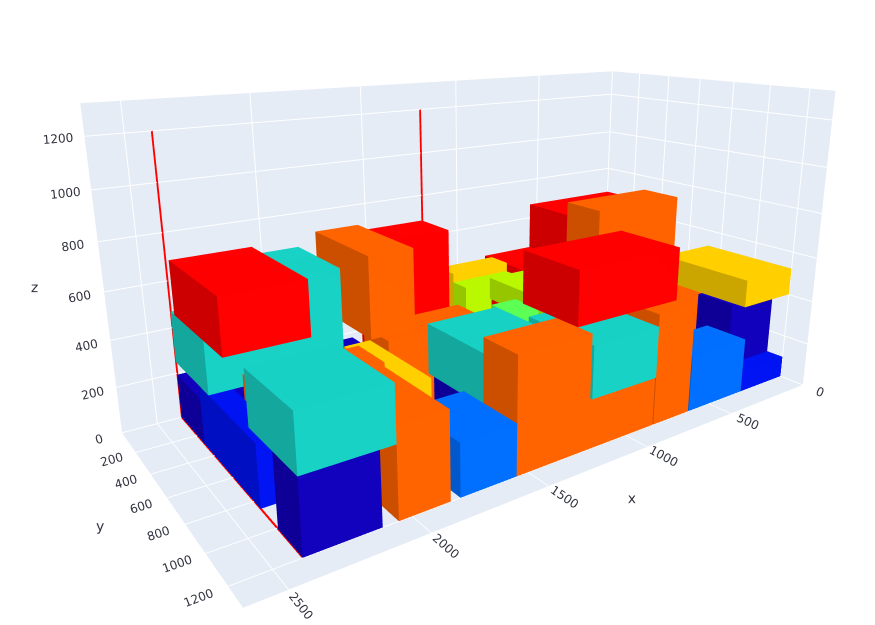}}\hspace{2mm}%
	\subfigure[\texttt{3dBPP\_3}: 52 items without restrictions.\label{fig:3dBPP_2}]{\includegraphics[width=\linewidth/4-1.5mm]{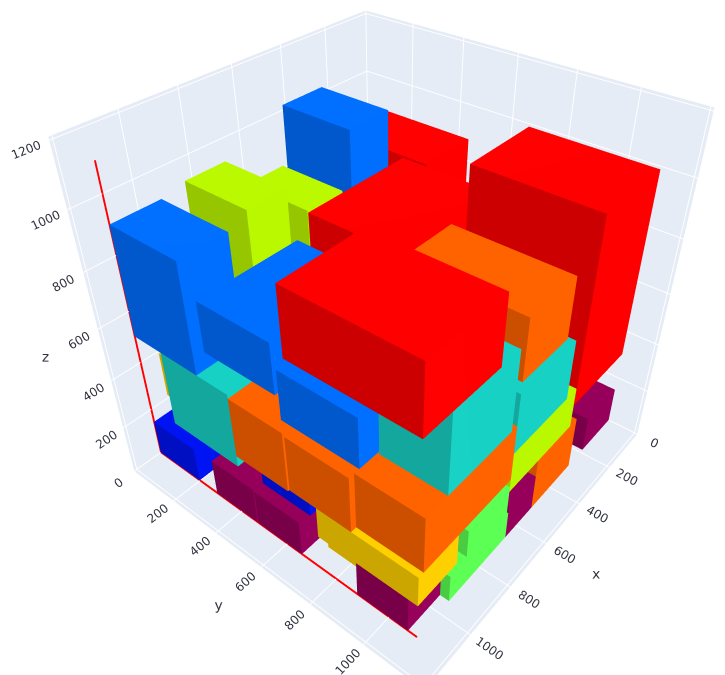}}\hspace{2mm}%
	\subfigure[\texttt{3dBPP\_4}: inst. \texttt{3dBPP\_3} with $\eta=2$. $\{\text{0:} {\crule[Wine]}, \text{1:} {\crule[DarkBlue]},\text{9:} {\crule[Red]}\}$ weight more than twice of $\{\text{3:} {\crule[LightBlue]},\dots,\text{8:} {\crule[Orange]}\}$.\label{fig:3dBPP_2_fra}]{\includegraphics[width=\linewidth/4-1.5mm]{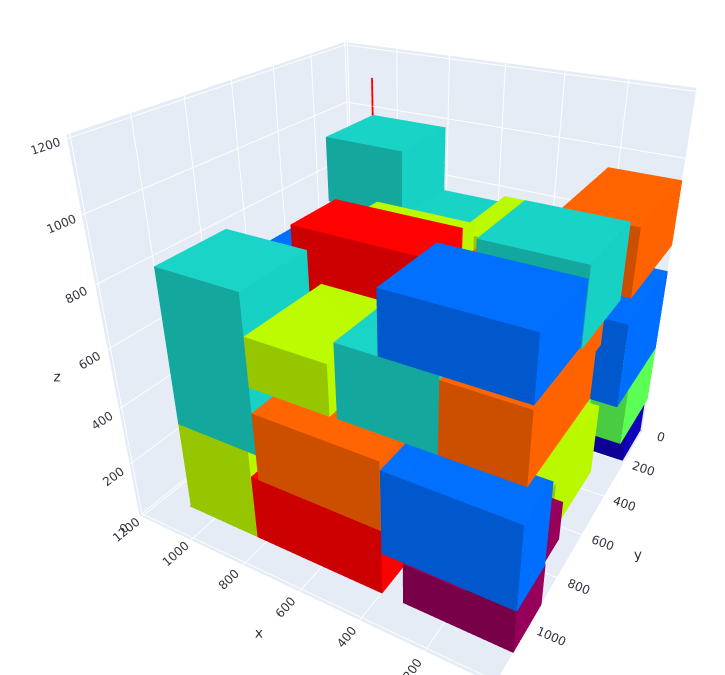}}
    \subfigure[\texttt{3dBPP\_5}: 54 items without restrictions.\label{fig:3dBPP_3}]{\includegraphics[width=\linewidth/4-1.5mm]{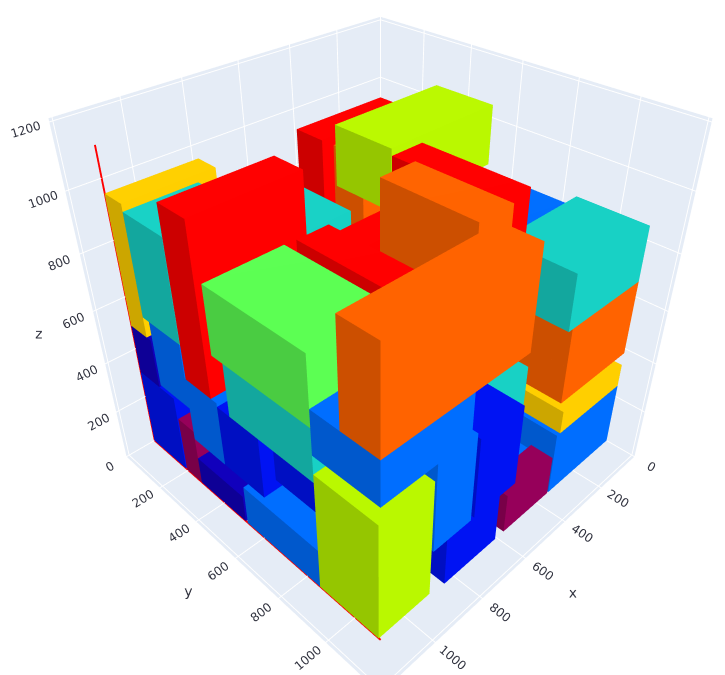}}\hspace{2mm}%
	\subfigure[\texttt{3dBPP\_6}: inst. \texttt{3dBPP\_5} with $\{(\text{4:} {\crule[Teal]}, \text{7:} {\crule[Yellow]}),(\text{7:} {\crule[Yellow]}, \text{9:} {\crule[Red]})\}$ incompatible.\label{fig:3dBPP_3_aff}]{\includegraphics[width=\linewidth/4-1.5mm]{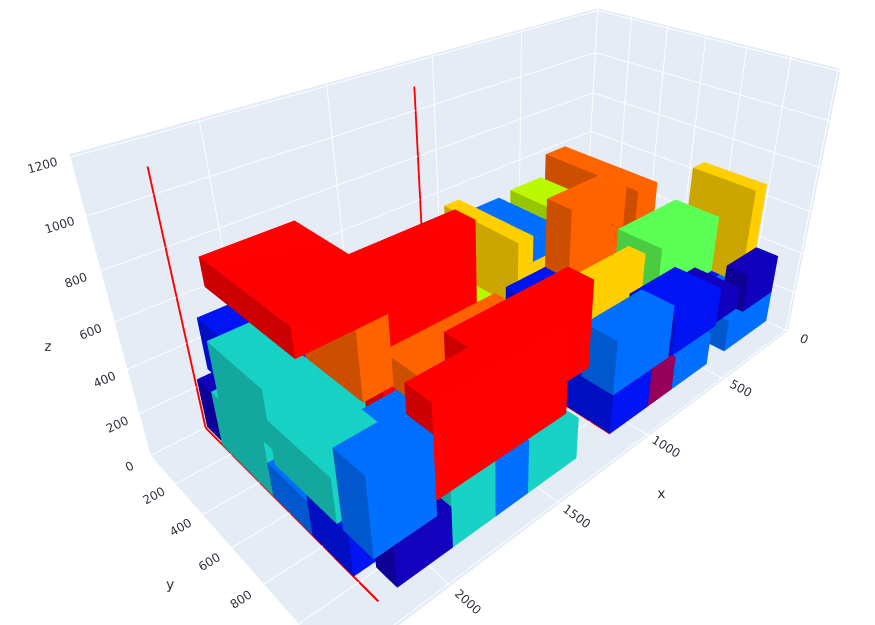}}\hspace{2mm}%
	\subfigure[\texttt{3dBPP\_7}: 46 items without restrictions.\label{fig:3dBPP_4}]{\includegraphics[width=\linewidth/4-1.5mm]{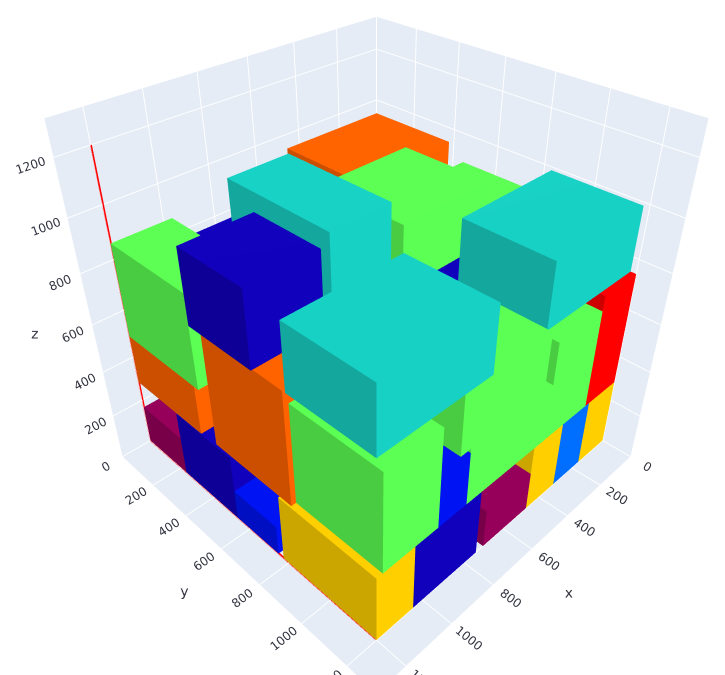}}\hspace{2mm}%
	\subfigure[\texttt{3dBPP\_8}: inst. \texttt{3dBPP\_7} with $(\text{4:} {\crule[Teal]}, \text{8:} {\crule[Orange]})$ incompatible and $\{(\text{0:} {\crule[Wine]}, \text{3:} {\crule[LightBlue]}),(\text{0:} {\crule[Wine]}, \text{8:} {\crule[Orange]})\}$ together.\label{fig:3dBPP_4_aff}]{\includegraphics[width=\linewidth/4-1.5mm]{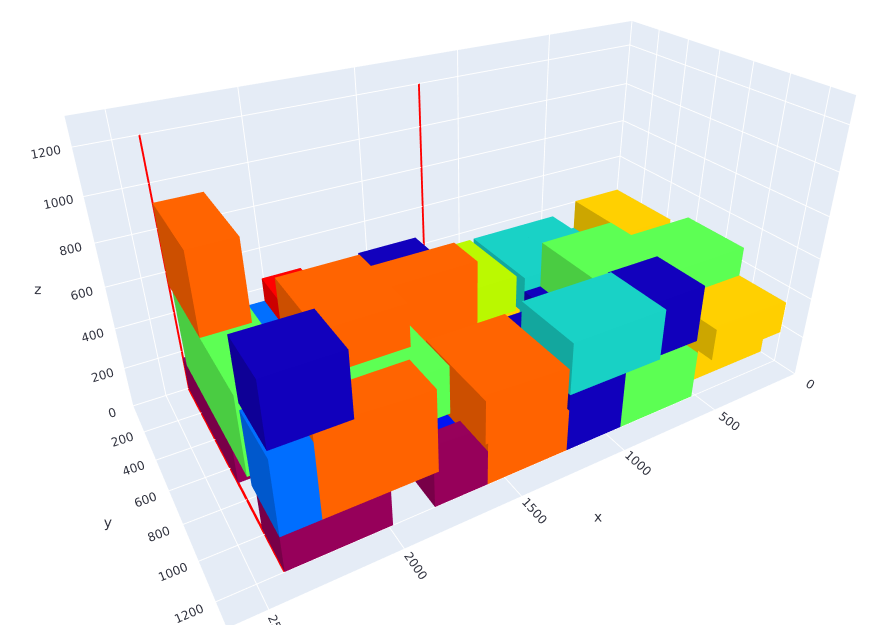}}
    \subfigure[\texttt{3dBPP\_9}: center of mass at $(\tilde{L},\tilde{W})=(750,750)$, the middle of the bin.\label{fig:3dBPP_5_CM1}]{\includegraphics[width=\linewidth/4-1.5mm]{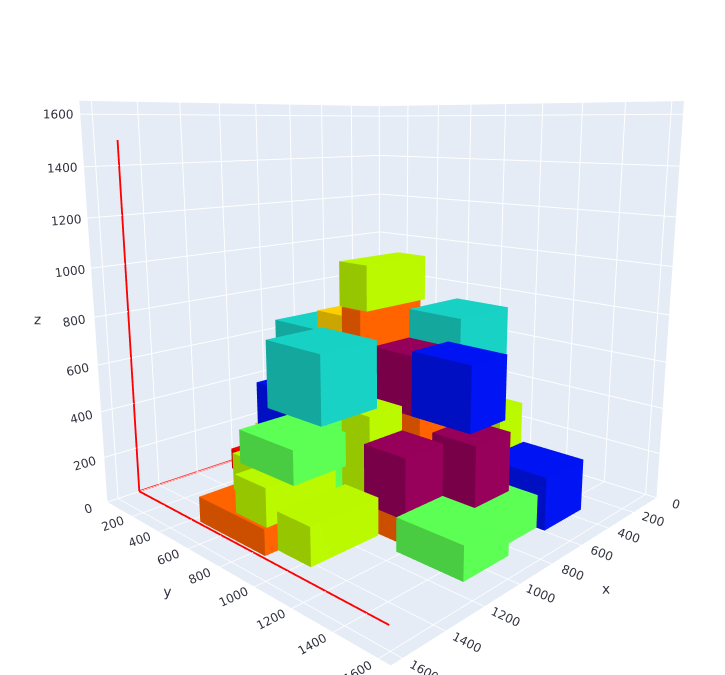}}\hspace{2mm}%
	\subfigure[\texttt{3dBPP\_10}: center of mass at $(\tilde{L},\tilde{W})=(900,500)$.\label{fig:3dBPP_5_CM2}]{\includegraphics[width=\linewidth/4-1.5mm]{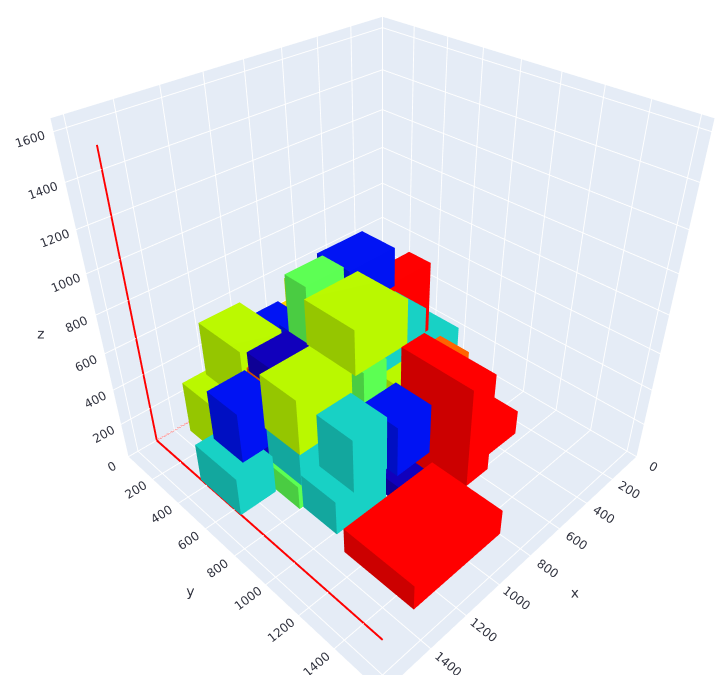}}\hspace{2mm}%
	\subfigure[\texttt{3dBPP\_11}: $M=800$; $\eta=2$ and $\{\text{0:} {\crule[Wine]},\text{7:} {\crule[Yellow]}\}$ weight more than twice of the rest; $(\text{7:} {\crule[Yellow]}, \text{9:}{\crule[Red]})$ incompatible; $\{(\text{0:}{\crule[Wine]},\text{3:}{\crule[LightBlue]}),(\text{0:}{\crule[Wine]},\text{8:}{\crule[Orange]})\}$ together; $(\tilde{L},\tilde{W})=(750,750)$.\label{fig:3dBPP_6_ALL1}]{\includegraphics[width=\linewidth/4-1.5mm]{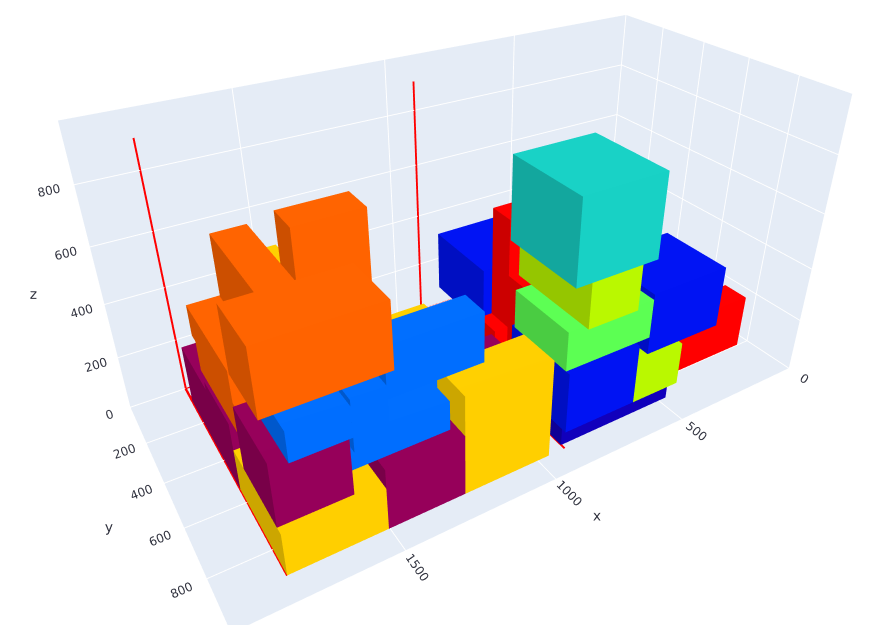}}\hspace{2mm}%
	\subfigure[\texttt{3dBPP\_12}: $M=900$; $\eta=2$ and $\{\text{3:}{\crule[LightBlue]},\text{4:}{\crule[Teal]}\}$ weight more than twice of the rest; $(\text{4:}{\crule[Teal]},\text{8:}{\crule[Orange]})$ incompatible; $(\text{2:}{\crule[Blue]},\text{4:}{\crule[Teal]})$ packed together; $(\tilde{L},\tilde{W}) = (500,500)$ (the middle).\label{fig:3dBPP_6_ALL2}]{\includegraphics[width=\linewidth/4-1.5mm]{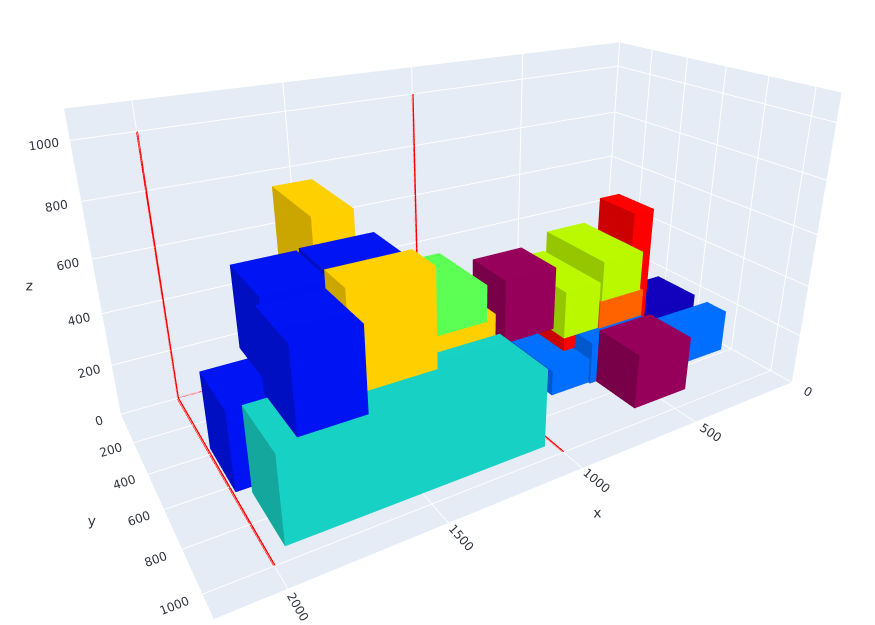}}\vspace{-3mm}%
	\caption{\textbf{(a)} Colour palette used in the solved instances. \textbf{(b)}-\textbf{(m)} Brief description of instances given in~\tablabel{tab:instances} and solutions provided by \texttt{Q4RealBPP}. Red lines show the bin boundaries. The activated restrictions work as expected.}\label{fig:results}
\end{figure}%

Besides that, we have conducted additional experimentation with the goal of further understanding the performance of \texttt{Q4RealBPP}. For this purpose, we have solved all the instances described in~\tablabel{tab:instances} under different time limits. Thus, each instance has been run 10 times with 5, 10, 30 and 60 seconds as time limits. In the top of~\figlabel{fig:times}, the obtained results are shown using the mean $\mu$ and and standard deviation $\sigma$ of the energy value $\{e_i\}_{i=1}^{10}$, being $e_i$ the $i$-th energy returned by the solver upon the assigned time limit for each instance. Moreover, the mean QPU access time in nanoseconds is also attached. As a complementary analysis, in the botton part of~\figlabel{fig:times}, since the returned energies vary considerably depending on the nature of the instance, we have computed the deviation around the mean energies in terms of
\begin{equation}\label{eq:dev_mean}
 \overline{\sigma}=\frac{1}{10}\sum_{i=1}^{10}\left|\frac{e_i}{\mu}-1\right|
\end{equation}
for illustrating more clearly the stability of the solution. This additional analysis is depicted in the bottom part of \figlabel{fig:times}.

Three main conclusions can be drawn from \figlabel{fig:times}: \textit{1)} generally, the longer the time limit given, the lower the energy returned (thus the better solution quality); \textit{2)} the deviation around the mean values of the objective function remains stable against the different time limits given as well as it does not show a remarkable dependency with the nature of the instance studied; and \textit{3)} although different time limits have been given to the solver, the QPU access time did not vary significantly for each one, indicating that the optimization process on the \texttt{LeapCQMHybrid} workflow mainly relies on the heuristic module of the solver. Despite the clear behaviour exhibited by our results, further work should be done for having a deeper understanding of how the presented framework performs.

\noindent\begin{figure}[!tbp]
\centering
\includegraphics[width=\linewidth]{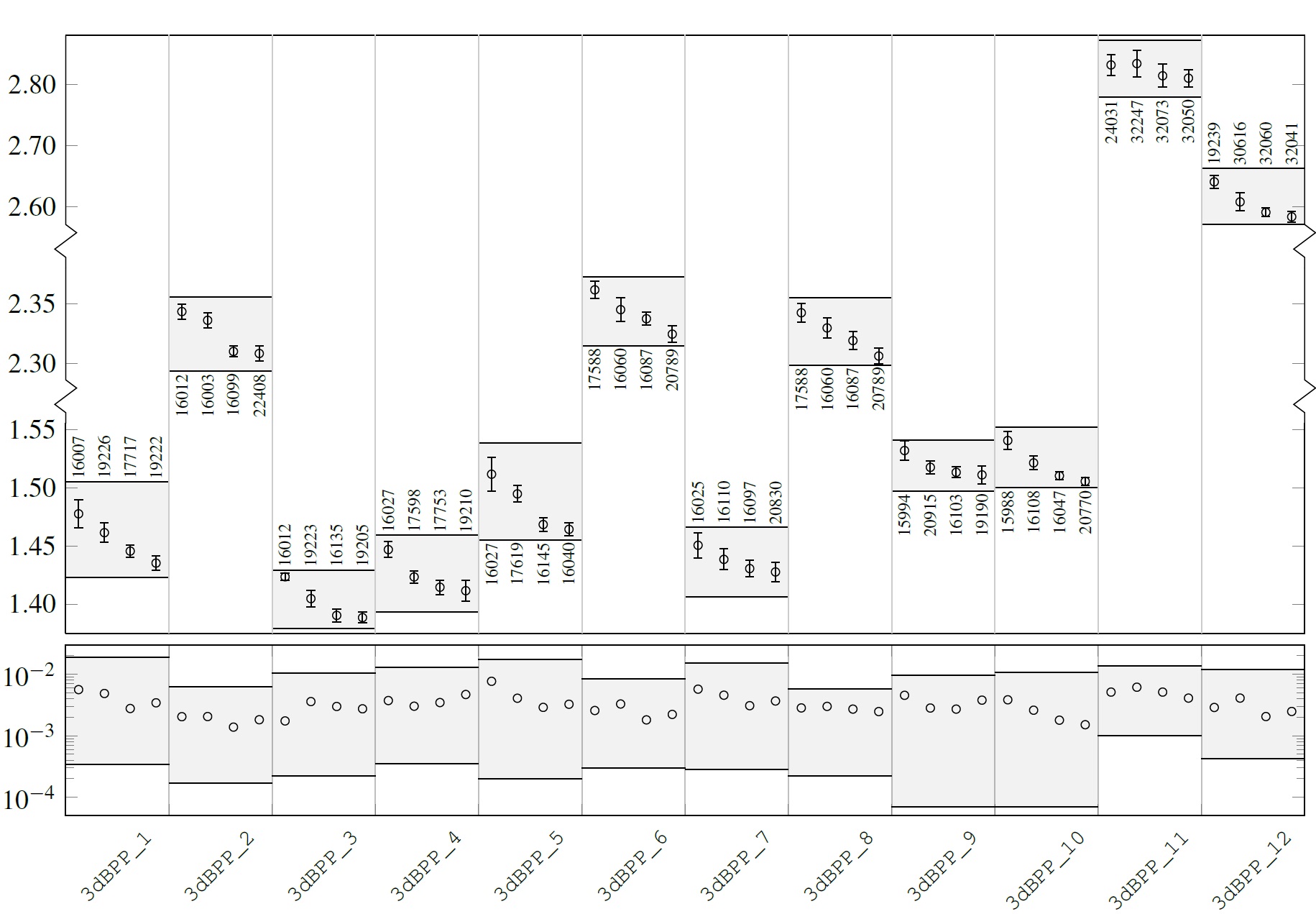}%
\caption{Top: mean + std for the 10 runs belonging to each instance. From left to right on each instance, the time limits given to the solver  are $\SI{5}{s}$, $\SI{10}{s}$, $\SI{30}{s}$ and $\SI{60}{s}$. Numbers placed above and below the shaded areas show the mean QPU access time for every time limit within each instance (in units of nanoseconds). Bottom: using~\eqref{eq:dev_mean}, deviation around the mean energy of the results depicted above. For both plots, shaded area shows the region where the minimum and maximum values fell during the study.}\label{fig:times}
\end{figure}%

\starsection{Conclusions and future work}\label{sec:conc}

In this work we have presented \texttt{Q4RealBPP}, a quantum-classical framework for solving real-world instances of the 3dBPP. This software can be easily used for both end-users and practitioners for dealing with 3dBPP instances considering constraints such as the weight, load bearing, package categories and load balancing.

To prove the applicability of \texttt{Q4RealBPP}, we have tested it over 12 instances of different nature, with the main intention of showcasing the capacity of the method to encompass real-world constraints. As depicted in~\figlabel{fig:results}, \texttt{Q4RealBPP} has successfully tackled all the generated instances, contemplating different real-world situations. Particularly noteworthy are the last two instances, \texttt{3dBPP\_11} and \texttt{3dBPP\_12} (Figures~\ref{fig:3dBPP_6_ALL1} and~\ref{fig:3dBPP_6_ALL2}, respectively), where all the constraints are activated.

The future work comprises 3 main interests: \textit{i)} to develop a more advanced version of the \texttt{Q4RealBBP} to generalise the framework to other BPP variants, and considering further features such as multi-class categorisation of items; \textit{ii)} to exploit this framework in conjunction to complementary Artificial Intelligence algorithms to enhance its potential real applications; and iii) to further analyse the results by comparing performance with classical solvers.

\starsection{Data availability}\label{sec:data}

The code is available from the corresponding author (E.O.) upon reasonable request. The data used, as well as \texttt{Q4RealBPP- DataGen} and all the results discussed in this work, are available at \url{http://dx.doi.org/10.17632/y258s6d939.1}.

\bibliography{sample}

\starsection{Funding}

This work was supported by the Basque Government through HAZITEK program (Q4\_Real project, ZE-2022/00033), and by the Spanish CDTI through Proyectos I+D Cervera 2021 Program (QOptimiza project, 095359) and Misiones Ciencia e Innovación Program (CUCO) under Grant MIG-20211005.

\starsection{Author contributions statement}

All authors conceived the research. S.V.R. formulated the problem and developed the code. E.V.R. formulated and developed the instance generator. S.V.R. and E.O. conceived and conducted the experiments. All authors wrote the manuscript. All authors reviewed the manuscript.

\starsection{Competing interests}

The authors declare no competing interests.

\end{document}